\documentstyle[twocolumn,psfig]{l-aa}

\def\kms{$\rm km\;s^{-1}$}
\def\arcsec{$''$}
\def\arcmin{$'$}

\def\ha{H$\alpha$}

\def\nii{[N~{\scriptsize II}]}

\begin{document}

\thesaurus{03(11.09.1 UGC~10205; 
 	      11.11.1;           
	      11.19.2;           
	      11.19.6)}          

\title{
Figure-of-eight velocity curves: UGC~10205\thanks{Based
on observations carried out at the INT operated on the island La Palma by the
Royal Greenwich Observatory in the Spanish Observatorio del Roque de Los
Muchachos of the Instituto de Astrof\'{\i}sica de Canarias, Tenerife, Spain.} }

\author{J.C.~Vega\inst{1},
	E.M.~Corsini\inst{2},
	A.~Pizzella\inst{2},
	and F.~Bertola\inst{2} }
\offprints{J.C. Vega; {\tt jvega@astrpd.pd.astro.it}}
 
\institute{Telescopio Nazionale Galileo, Osservatorio Astronomico di Padova, 
vicolo dell'Osservatorio 5, I-35122 Padova, Italy
\and Dipartimento di Astronomia, Universit{\`a} di Padova,
vicolo dell'Osservatorio 5, I-35122 Padova, Italy }

\date{Received..................; accepted...................}

\maketitle

\markboth{J.C. Vega et al.: Figure-of-eight velocity curves: UGC~10205}
{J.C. Vega et al.: Figure-of-eight velocity curves: UGC~10205}

\begin{abstract}

We measured the velocity curve and the velocity dispersion profile
of the ionized gas along the major axis of the edge-on galaxy UGC~10205.  
The observed kinematics extends up to about 40\arcsec\ from the nucleus.
In the inner $\pm 13$\arcsec\ of this early-type spiral three 
kinematically distinct gaseous components are present.  
We disentangle a fast-rotating and a slow-rotating component. 
They give to the UGC~10205 velocity curve a ``figure-of-eight'' appearance.
A third velocity component is also detected on the southeast side of the galaxy. Possibly it is produced by gas in non-circular motions.


\end{abstract}

\section{Introduction}

Within the last year ionized gas kinematics has revealed in a number of 
edge-on disk galaxies double-peaked emission lines.  
The analysis of the line profiles allows to derive individual 
rotation curves characteristic of two kinematically distinct gas components.

The edge-on S0 galaxy NGC~7332 has been found to have an ``x-shaped'' velocity
curve, indicating two gas components counterrotating one with respect to the other (Plana \& Boulesteix 1996).\\ 
In the inner regions of the Sc NGC~5907 Miller \& Rubin (1995) observed
double-valued ionized gas emissions. They attributed the higher velocity
system to disk gas near the nucleus, and the lower velocity system to an
outer gas ring. Although these two gas components are supposed to be 
spatially distinct, they are viewed superimposed along the line-of-sight 
on account of NGC~5907 high inclination.\\   
The spirals NGC~5746 (Kuijken \& Merrifield 1995; Bureau \& Freeman 1996), 
NGC~5965 (Kuijken \& Merrifield 1995), IC~5096 (Bureau \& Freeman 1996) and
NGC~2683 (Merrifield 1996) have boxy/peanut bulges and double-peaked
gas line profiles. 
Kuijken \& Merrifield (1995) explained these features as the signature of a 
disk non-axisymmetric potential due to the presence of a bar.

In this paper we show yet an other case of edge-on disk galaxy with a
multiple-valued gas velocity curve, namely UGC~10205. 
For projected distances lower than 13\arcsec\ from the nucleus
UGC~10205 it is characterized by the presence of three kinematically
distinct gaseous components, two of which give to its velocity curve a
``figure-of-eight '' appearance.\\ 
UGC~10205 is classified as Sa spiral by Nilson (1973) and by de Vaucouleurs et
al. (1991). Its total $B$-band magnitude is $B_T=14.4$ mag (RC3) and the
inclination of the galaxy deduced from the disk parameters is $i=84^\circ$
(Rubin et al. 1985). The distance is 132 Mpc ($H_0 = 50$ \kms\ Mpc$^{-1}$).

\section{Observations and data reduction}

\begin{figure*}[ht]
\centerline{{\psfig{figure=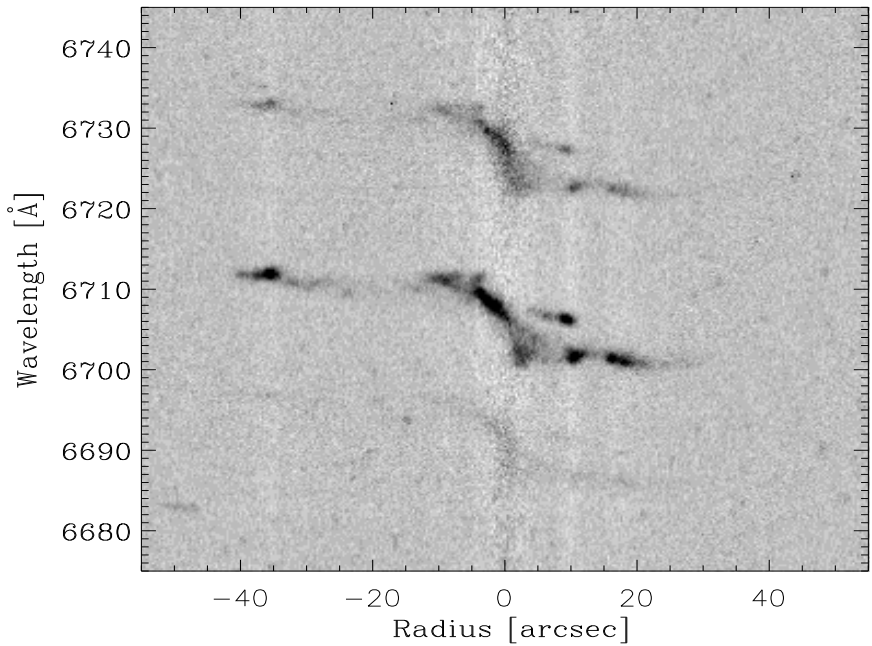,height=75mm,width=120mm}} }
\vspace*{.5cm}
\caption[Figura 1]{UGC~10205 major axis 100 minutes spectrum in the \ha\ region 
after removal of sky and stellar continuum. Note 
the multiple-peaked emissions of the \ha\ and the \nii\
($\lambda\,6583.4$ \AA) lines in the central regions 
of the spectrum ($|r| \leq 13$\arcsec ) }
\end{figure*}

The spectroscopic observations of  UGC~10205 were carried out on March 19-21,
1996 at the Isaac Newton Telescope (INT) in La Palma using the Intermediate
Dispersion Spectrograph (IDS).

The H1800V grating with 1800 grooves$\rm\;mm^{-1}$ was used in the first order
in combination with a 1.9\arcsec\ $\times$ 4.0\arcmin\ slit, the 500~mm camera
and the AgRed collimator. It yielded a wavelength coverage of $\sim240$ \AA\
between 6650~\AA\ and 6890~\AA\ with a reciprocal dispersion of 
9.92~$\rm\AA\;mm^{-1}$.  
 We checked that the measured
FWHMs do not depend on wavelength and we found a mean value of FWHM = 0.86~\AA\
(i.e.  $\sigma = 0.37$~\AA) that, in the range of the observed gas emission
lines, corresponds to $\sim17$~\kms .
No on-chip binning was done on the adopted 1024$\times$1024 TK1024A CCD. Each 
24~$\mu$m~$\times$~24~$\mu$m image pixel corresponds to
0.24~\AA~$\times$~0.33\arcsec .

We took two separate major axis spectra (P.A. = 132$^\circ$) for a total
exposure time of 100 minutes (Fig.~1).
The slit was centered visually on the galaxy nucleus. A comparison copper-argon lamp exposure was obtained between the two object integrations.
  All the images were reduced using standard MIDAS routines.
Considering a sample of 8 bright  OH night-sky emission lines, we found a mean deviation
from the theorical predicted wavelength (Osterbrock \& Martel, 1992) corresponding to $\sim 1$~\kms. 

The gas velocities and velocity dispersions  for $|r| > 13$\arcsec, where the
ionized gas emission lines have a Gaussian profile shape, were derived by means
of the MIDAS package ALICE.  We measured the \ha\ and the \nii\
($\lambda\,6583.4$ \AA) lines, where they were clearly detected. The position, 
the FWHM and the uncalibrated flux of each emission line were individually
determined by interactively fitting one Gaussian plus a polynomial to each
emission and to its surrounding continuum. The wavelength of the Gaussian center
was converted to the velocity $v = cz$, and then an heliocentric correction of
$\Delta v=+15.5$ \kms\ was applied. The Gaussian FWHM was corrected for 
the instrumental FWHM and then converted to the velocity dispersion $\sigma$.
The ionized gas emission lines are double-peaked for $-13$\arcsec\
$\leq r <$ 3\arcsec\ and even triple-peaked for 3\arcsec\ $\leq r \leq 13$\arcsec .
They have been fitted  using the above package with two
or three Gaussians respectively and a polynomial continuum. At each radius this
multiple-Gaussian fit has been done separately for each emission line.

The gas velocity curves and the velocity dispersion profiles independently
derived from the \ha\ (Fig.~2) and the \nii\ (Fig.~3) lines are in good
agreement at all radii.
The kinematical data from
\ha\ and \nii\ lines are given in Table~1 and in Table~2 respectively. Each
table provides the radial distance from the galaxy center $r$ in arcsec
(col.~1), the observed heliocentric velocity $v$ (col.~2) and the velocity
dispersion $\sigma$ (col.~3) in \kms , the number n of spectrum rows binned
along the spatial direction to improve the signal-to-noise ratio of the emission
lines (col.~4) and the identification i of the kinematically distinct 
gas components (col.~5).

\section{Results}

\begin{figure*}
\centerline{{\psfig{figure=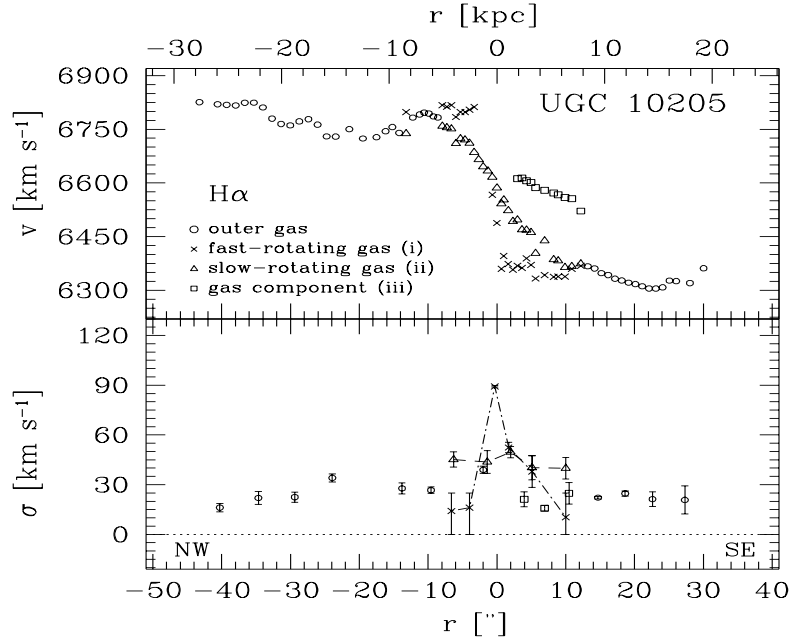,height=95mm,width=120mm}} }
\caption[Figura 2]{ UGC~10205 major-axis \ha\ kinematics: observed heliocentric
velocity curve (top) and velocity dispersion profile (bottom) of the gaseous
fast-rotating ({\em crosses\/}), slow-rotating ({\em open triangles\/}), and
third ({\em open squares\/}) component. The {\it open circles\/} are
used where \ha\ was fitted using a single Gaussian. 
The velocity dispersion $\sigma$ is plotted with a different spatial binning than $v$.
The {\it dash-dotted\/} and the {\it short-dashed lines\/} connect $\sigma$ values of the fast and the slow-rotating components respectively }
\end{figure*}

The observed gas kinematics extends out to 42\arcsec\ ($\sim27$~kpc) in 
the receding NW side and about up to 30\arcsec\ ($\sim19$~kpc) in the SE
approaching side respectively.

For $|r| \leq 13$\arcsec\ ($\sim8$ kpc) we are able to disentangle 
(Fig.~1 and Fig.~2) three kinematically distinct gaseous components, named 
as fast-rotating (i), slow-rotating (ii) and third (iii).\\ 
The fast-rotating gas component (i) shows a velocity curve with a very steep
gradient, reaching an observed maximum rotation of 240~\kms\ at 
$|r|\sim3$\arcsec\ ($\sim2$~kpc) from the center and remaining almost 
constant for $|r| > 3$\arcsec .  
Its velocity dispersion has a central peak of about 90~\kms\ and it shows a
sharp decrease to values lower than 25~\kms\ outwards. The radial velocity of
the slow-rotating gas component (ii) increases linearly with the distance from
the galaxy center reaching 240~\kms\ at $|r|\sim14$\arcsec\ ($\sim9$~kpc). The
velocity dispersion remains between 40~\kms\ and 50~\kms .  
This range is larger in the \nii\ line, which however is characterized by a
lower signal-to-noise ratio.
In the radial range between $r\sim3$\arcsec\ and $r\sim13$\arcsec\ along the 
SE side of the major axis the \ha\ and \nii\ emissions have triple-peaked lines.
Indeed in this region the component (iii) is observed. It has a radial 
velocity increasing linearly from $v\sim6551$~\kms\ to $v\sim6611$~\kms\ 
and equal to the systemic velocity at $r \sim 8$\arcsec\ ($\sim5$ kpc).
It has a quite low velocity dispersion of $\sigma\sim15$ \kms .\\
For $|r| > 13$\arcsec\ a single-valued velocity curve is measured,
showing the tendency to flatten out. 
The velocity dispersions in this radial range are lower than 30~\kms .

\begin{figure*}
\centerline{{\psfig{figure=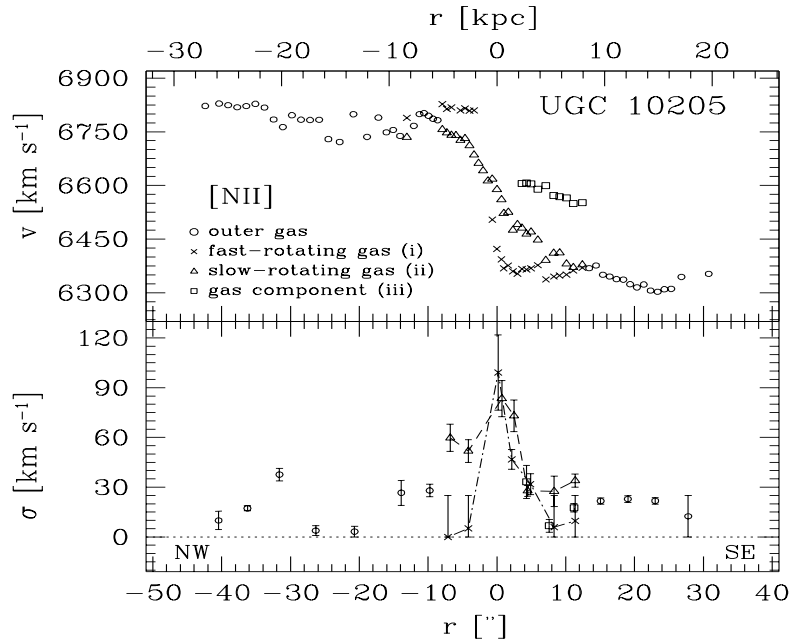,height=95mm,width=120mm}} }
\caption[Figura 2]{Same as Fig.~2 for the \nii\ ($\lambda\,6583.4$ \AA) emission line }
\end{figure*}

Adopting the center of symmetry of velocities for $|r| \leq 13$\arcsec\
as the systemic heliocentric velocity, we derived $V_\odot = 6583 \pm 5$~\kms\
in agreement with $V_\odot = 6581 \pm 15$~\kms\ found by Rubin et al. (1985).

Studying the ionized gas velocity curves of a sample of Sa spirals, also Rubin
et al. (1985) noticed along the SE side of the UGC~10205  major axis ``a curious
three-velocity system'' within 10 kpc of the center. It can not be reconciled  
with our component (i), (ii), and (iii).
The reciprocal dispersion (25 $\rm\AA\;mm^{-1}$) and the spatial scale
(25\arcsec\ mm$^{-1}$) of their image-tube spectrum were respectively 3 and 2
times lower than those of our CCD spectrum. 
So they did not disentangle the component (i) from the (ii), detecting
them as a unique one. They observed a second velocity system,
corresponding to velocity curve of  component (iii). Finally, for two distinct
radii at $r\sim -5$\arcsec\ they measured intermediate velocities between those
of the first and the second system and considered them as related to a third
velocity component.

\section{Discussion}

In the inner $\pm 13$\arcsec\ UGC~10205 we are facing two main kinematically distinct gaseous components, namely (i) and (ii). 
They have quite similar velocity dispersion profiles but
very different velocity curves, which produce the ``figure-of-eight'' 
appearance of UGC~10205 velocity curve.
  
What is the real spatial distribution of these two components? Are they really
cospatial or are they spatially distinct and seen superimposed on account of a
projection effect?

The simultaneous presence of the two gas components at the same distance of the
galaxy center raises the problem of the viscous interaction of distinct gaseous
structures with different kinematical characteristics.  One possibility is the
gas to be distributed in collisionless cloudlets, as suggested by Cinzano \& van
der Marel (1994) to explain the ionized gas kinematics in the E4 NGC~2974. 
If this is the case, we would expect the slow-rotating component being supported
by a velocity dispersion higher than that we observed. 

We are left with the interpretation already given by Miller \& Rubin (1995) for
NGC~5907 and by Kuijken \& Merrifield (1995) for NGC~5746 and NGC~5965 that the
two gas components are spatially distinct and viewed superimposed along the
line-of-sight due to the high inclination of the galaxy. The linear rise of the
velocity curve of component (ii) up to the points of conjunction with those of
component (i) is due to the so-called ``rim of the wheel'' effect
when viewing at an enhanced ring structure. 
Moreover, the radial trend of line intensity derived for the two components
seem to confirm this interpretation. Indeed, the intensity of component (i) 
is peaked in the center, while the intensity of component (ii) is almost 
constant as we are expecting if component (ii) is actually an outer ring.
  
The above gas configuration is the one expected in a barred galaxy.
Indeed the bar exerts a torque on the disk gas, which is slowly drifted 
from the regions around the corotation radius towards the Lindblad resonances 
to form rings. 
Kuijken \& Merrifield (1995) and Merrifield (1996) showed the
line-of-sight velocity distribution (LOSVD) in function of the projected 
radius for the closed non-intersecting orbits allowed by a barred 
disk potential in edge-on galaxies. 
(Due to its collisional nature, the gas moves only onto the closed
 non-intersecting orbits.)
They found the LOSVDs with the characteristic ``figure-of-eight'' variation 
with radius. 
For this reason they considered the gaps in such LOSVDs as the signature of the presence in the disk of the gas-depleted regions due to the bar. 
The gas components (ii) in the velocity curve of UGC~10205 is
produced by the ring formed at the outer Lindblad resonance.

The third gas component observed in the inner regions of UGC~10205 
is very peculiar. It is present only on the SE side and it has a velocity
ranging from $\sim-30$ \kms\ to $\sim+30$ \kms\ if reported to the systemic velocity of the galaxy. 
So it is not moving around the galaxy center in circular orbits.
It could be associated to the faint features embedding UGC~10205 and
visible in the $R$-band images shown by Rubin (1987).
Its kinematics can be explained if such gas moves onto an elliptical orbit,
which at the projected distance of $r\sim8$\arcsec\ has its tangent
perpendicular to the line-of-sight.
The gas of component (iii) populates only a portion of this orbit.
From the available data we can not infer the proper distance of component (iii)
from the galaxy center.

Outside 13\arcsec\ we observe single-peaked emission lines produced 
by disk gas in near-circular motion. For radial distances greater than that 
of the outer Lindblad resonance we expect the gas to be only little disturbed
by the inner triaxial potential.
Because of the edge-on orientation of the disk with respect to the 
line-of-sight, if the gas is distributed uniformly decreasing throughout the disk
we would expect to observe emission lines peaked at the local circular velocity 
with an asymmetry towards the lower velocities.
This emission feature is present in UGC 10205 in the form of very low 
luminous intensity. It is superimposed to a major emission that does not show any systematic deviation from the Gaussian shape. 
Irregularities in the velocity curve can be easily produced by the fact of
the integration along the line-of-sight being the galaxy seen on edge. 
   
The interpretation of Kuijken \& Merrifield (1995) has been recently 
adopted by Merrifield (1996) for NGC~2683 observed by Rubin, and by
Bureau \& Freeman (1996) for IC~5096. It could also be applied not only to
UGC~10205 but also to the case NGC~5907, extending the explanation given by
Miller \& Rubin (1995).  
Since their two spirals have peanut-shaped bulges, Merrifield and Kuijken 
(1995) suggested a connection between the peanut bulges, which are
detectable only in edge-on galaxies, and the bars, which are easily 
detectable in more face-on systems. 
Anyway, while NGC~2683, NGC~5746, NGC~5965 and IC~5096
have boxy/peanut-shape bulges, for NGC~5907 and UGC~10205 this crucial
photometric information is still not available.
 
The presence of the ``figure-of-eight'' in gas velocity curves of edge-on
spirals seems to be unrelated to the morphological type in the spiral 
sequence since in RC3 UGC~10205 is classified Sa, NGC~2683, NGC~5746 and
NGC~5965 are Sb, IC~5096 is Sbc, and NGC~5907 is Sc.
  
The geometrical distribution of the gas giving rise to the ``figure-of-eight''
velocity curves could be used in the interpretation of phenomena like the
one described by Plana \& Boulesteix (1996) in the S0 NGC~7332.
It is an edge-on galaxy with a boxy-shaped bulge (Fisher \& Illinghworth 1994).
Plana \& Boulesteix (1996) separated clearly two extended coplanar
counterrotating components of ionized gas, which were previously detected 
by Fisher \& Illinghworth (1994).
One of these components has a linearly increasing velocity curve (``rim of
the wheel'' effect). 
If we imagine to invert the sense of rotation of this latter component,
a typical ``figure-of-eight'' velocity curve is obtained.
Therefore a gas distribution like that of the above-mentioned edge-on barred
spirals but with an outer ring counterrotating with respect to the inner gas
could reproduce the kinematics of NGC~7332, for which a satisfactory
interpretation has been not given.
Of course, theoretical modelling should justify this situation.

\begin{acknowledgements}
We thank M.R. Merrifield for useful discussion.
AP acknowledges support from an {\it Acciaierie Beltrame\/} grant.
\end{acknowledgements}


\clearpage
\begin{table}[t]
\caption{UGC~10205 major axis \ha\ kinematics}
\begin{flushleft}
\begin{scriptsize}
\begin{tabular}{lllllllllll}
\hline   {}
 & & & & & & & & & & \\
\multicolumn{1}{c}{\normalsize{$r$}} &    
\multicolumn{1}{c}{\normalsize{$v$}} & 
\multicolumn{1}{c}{\normalsize{$\sigma$}} & 
\multicolumn{1}{c}{\normalsize{n}} & 
\multicolumn{1}{c}{\normalsize{i$^{\rm a}$}} &
\multicolumn{1}{c}{\hspace{.1truecm}} &
\multicolumn{1}{c}{\normalsize{$r$}} &    
\multicolumn{1}{c}{\normalsize{$v$}} & 
\multicolumn{1}{c}{\normalsize{$\sigma$}} & 
\multicolumn{1}{c}{\normalsize{n}} & 
\multicolumn{1}{c}{\normalsize{i$^{\rm a}$}} \\ 
\multicolumn{1}{c}{(1)} & 
\multicolumn{1}{c}{(2)} & 
\multicolumn{1}{c}{(3)} & 
\multicolumn{1}{c}{(4)} & 
\multicolumn{1}{c}{(5)} &
\multicolumn{1}{c}{\hspace{.1truecm}} &
\multicolumn{1}{c}{(1)} & 
\multicolumn{1}{c}{(2)} & 
\multicolumn{1}{c}{(3)} & 
\multicolumn{1}{c}{(4)} & 
\multicolumn{1}{c}{(5)} \\
 & & & & & & & & & & \\
\cline{1-5}\cline{7-11}
 & & & & & & & & & & \\
$-43.2$ &  6826 & 23 &8 &4 & & $+  1.2$ &  6553 & 59 &1 &2 \\
$-40.6$ &  6820 & 12 &4 &4 & & $+  1.7$ &  6373 & 56 &2 &1 \\
$-39.3$ &  6818 & 13 &4 &4 & & $+  1.7$ &  6522 & 51 &2 &2 \\
$-38.0$ &  6817 & 18 &4 &4 & & $+  2.3$ &  6358 & 44 &2 &1 \\
$-36.6$ &  6824 & 18 &4 &4 & & $+  2.3$ &  6493 & 48 &2 &2 \\
$-35.3$ &  6824 & 20 &4 &4 & & $+  3.0$ &  6369 & 51 &2 &1 \\
$-34.0$ &  6811 & 33 &4 &4 & & $+  3.0$ &  6497 & 42 &2 &2 \\
$-32.7$ &  6780 & 16 &4 &4 & & $+  3.0$ &  6612 & 17 &2 &3 \\
$-31.4$ &  6765 & 21 &4 &4 & & $+  3.6$ &  6363 & 32 &2 &1 \\
$-30.0$ &  6761 & 28 &4 &4 & & $+  3.6$ &  6469 & 45 &2 &2 \\
$-28.7$ &  6772 & 14 &4 &4 & & $+  3.6$ &  6614 & 13 &2 &3 \\
$-27.4$ &  6778 & 26 &4 &4 & & $+  4.3$ &  6389 & 64 &2 &1 \\
$-26.1$ &  6763 & 28 &4 &4 & & $+  4.3$ &  6468 & 22 &2 &2 \\
$-24.8$ &  6730 & 36 &4 &4 & & $+  4.3$ &  6606 & 34 &2 &3 \\
$-23.4$ &  6730 & 40 &6 &4 & & $+  5.0$ &  6371 & 40 &2 &1 \\
$-21.5$ &  6750 & 33 &6 &4 & & $+  5.0$ &  6462 & 29 &2 &2 \\
$-19.5$ &  6724 &  0 &6 &4 & & $+  5.0$ &  6601 & 21 &2 &3 \\
$-17.5$ &  6728 & 86 &4 &4 & & $+  5.6$ &  6333 &  6 &4 &1 \\
$-16.2$ &  6744 & 27 &3 &4 & & $+  5.6$ &  6403 & 64 &4 &2 \\
$-15.2$ &  6756 & 26 &3 &4 & & $+  5.6$ &  6587 & 13 &4 &3 \\
$-14.2$ &  6740 & 19 &3 &4 & & $+  6.9$ &  6342 & 47 &3 &1 \\
$-13.2$ &  6738 & 11 &3 &2 & & $+  6.9$ &  6438 & 42 &3 &2 \\
$-13.2$ &  6798 & 17 &3 &1 & & $+  6.9$ &  6579 & 17 &3 &3 \\
$-12.2$ &  6783 & 40 &3 &4 & & $+  8.3$ &  6337 &  0 &2 &1 \\
$-11.2$ &  6791 & 28 &2 &4 & & $+  8.3$ &  6386 & 63 &2 &2 \\
$-10.6$ &  6796 & 25 &2 &4 & & $+  8.3$ &  6572 & 18 &2 &3 \\
$ -9.9$ &  6794 & 25 &2 &4 & & $+  9.0$ &  6338 & 12 &3 &1 \\
$ -9.2$ &  6786 & 25 &2 &4 & & $+  9.0$ &  6382 & 28 &3 &2 \\
$ -8.6$ &  6783 & 33 &2 &4 & & $+  9.0$ &  6567 & 16 &3 &3 \\
$ -7.9$ &  6759 & 35 &2 &2 & & $+  9.9$ &  6338 & 16 &3 &1 \\
$ -7.9$ &  6818 & 14 &2 &1 & & $+  9.9$ &  6364 & 27 &3 &2 \\
$ -7.3$ &  6755 & 54 &2 &2 & & $+  9.9$ &  6559 & 15 &3 &3 \\
$ -7.3$ &  6813 &  0 &2 &1 & & $+ 10.9$ &  6360 & 16 &3 &1 \\
$ -6.6$ &  6751 & 60 &2 &2 & & $+ 10.9$ &  6367 & 41 &3 &2 \\
$ -6.6$ &  6817 &  0 &2 &1 & & $+ 10.9$ &  6556 & 26 &3 &3 \\
$ -5.9$ &  6710 & 46 &2 &2 & & $+ 12.2$ &  6368 &  8 &3 &1 \\
$ -5.9$ &  6784 & 40 &2 &1 & & $+ 12.2$ &  6374 & 40 &3 &2 \\
$ -5.3$ &  6724 & 47 &2 &2 & & $+ 12.2$ &  6522 & 42 &3 &3 \\
$ -5.3$ &  6798 & 16 &2 &1 & & $+ 13.2$ &  6367 & 21 &3 &4 \\
$ -4.6$ &  6720 & 30 &2 &2 & & $+ 14.2$ &  6361 & 24 &3 &4 \\
$ -4.6$ &  6798 & 17 &2 &1 & & $+ 15.2$ &  6348 & 21 &3 &4 \\
$ -4.0$ &  6711 & 33 &2 &2 & & $+ 16.2$ &  6342 & 22 &3 &4 \\
$ -4.0$ &  6804 & 12 &2 &1 & & $+ 17.2$ &  6332 & 25 &3 &4 \\
$ -3.3$ &  6685 & 36 &2 &2 & & $+ 18.2$ &  6327 & 25 &3 &4 \\
$ -3.3$ &  6813 & 19 &2 &1 & & $+ 19.1$ &  6321 & 28 &3 &4 \\
$ -2.6$ &  6666 & 37 &2 &2 & & $+ 20.1$ &  6317 & 20 &3 &4 \\
$ -2.0$ &  6645 & 36 &2 &2 & & $+ 21.1$ &  6311 & 25 &3 &4 \\
$ -1.3$ &  6633 & 43 &2 &2 & & $+ 22.1$ &  6305 & 32 &3 &4 \\
$ -0.7$ &  6567 & 89 &2 &1 & & $+ 23.1$ &  6305 & 16 &3 &4 \\
$ -0.7$ &  6616 & 44 &2 &2 & & $+ 24.1$ &  6308 & 12 &3 &4 \\
$  0.0$ &  6488 & 90 &2 &1 & & $+ 25.1$ &  6327 & 30 &3 &4 \\
$  0.0$ &  6586 & 36 &2 &2 & & $+ 26.1$ &  6326 & 39 &6 &4 \\
$ +0.7$ &  6360 & 50 &2 &1 & & $+ 28.1$ &  6320 & 15 &6 &4 \\
$ +0.7$ &  6542 & 70 &2 &2 & & $+ 30.0$ &  6361 &  0 &6 &4 \\
$ +1.2$ &  6396 & 62 &1 &1 & &	        &       &    &  & \\
 & & & & & & & & & & \\
\hline
\end{tabular}
\label{tabHa}
\end{scriptsize}
\begin{list}{}{}
\item[$^{\rm a}$] Identification of the gas components: 1, 2, 3 indicate
kinematic data referring to component (i), (ii), and (iii) respectively; 4 is
used for the gas in the outer regions of the disk.
\end{list}
\end{flushleft}
\end{table}

\clearpage
\begin{table}[t]
\caption{UGC~10205 major axis [N II] kinematics}
\begin{flushleft}
\begin{scriptsize}
\begin{tabular}{lllllllllll}
\hline
 & & & & & & & & & & \\
\multicolumn{1}{c}{\normalsize{$r$}} &    
\multicolumn{1}{c}{\normalsize{$v$}} & 
\multicolumn{1}{c}{\normalsize{$\sigma$}} & 
\multicolumn{1}{c}{\normalsize{n}} & 
\multicolumn{1}{c}{\normalsize{i$^{\rm a}$}} &
\multicolumn{1}{c}{\hspace{.1truecm}} &
\multicolumn{1}{c}{\normalsize{$r$}} &    
\multicolumn{1}{c}{\normalsize{$v$}} & 
\multicolumn{1}{c}{\normalsize{$\sigma$}} & 
\multicolumn{1}{c}{\normalsize{n}} & 
\multicolumn{1}{c}{\normalsize{i$^{\rm a}$}} \\ 
\multicolumn{1}{c}{(1)} & 
\multicolumn{1}{c}{(2)} & 
\multicolumn{1}{c}{(3)} & 
\multicolumn{1}{c}{(4)} & 
\multicolumn{1}{c}{(5)} &
\multicolumn{1}{c}{\hspace{.1truecm}} &
\multicolumn{1}{c}{(1)} & 
\multicolumn{1}{c}{(2)} & 
\multicolumn{1}{c}{(3)} & 
\multicolumn{1}{c}{(4)} & 
\multicolumn{1}{c}{(5)} \\
 & & & & & & & & & & \\
\cline{1-5}\cline{7-11}
 & & & & & & & & & & \\
$-42.4$ &  6822 & 19 &8 &4 & & $+  1.2$ &  6368 & 46 &1 &1 \\
$-40.4$ &  6829 &  0 &4 &4 & & $+  1.2$ &  6522 &104 &1 &2 \\
$-39.1$ &  6824 & 11 &4 &4 & & $+  1.7$ &  6376 & 61 &2 &1 \\
$-37.7$ &  6819 & 14 &4 &4 & & $+  1.7$ &  6525 & 59 &2 &2 \\
$-36.4$ &  6822 & 18 &4 &4 & & $+  2.3$ &  6360 & 48 &2 &1 \\
$-35.1$ &  6828 & 20 &4 &4 & & $+  2.3$ &  6475 & 91 &2 &2 \\
$-33.8$ &  6818 & 39 &4 &4 & & $+  3.0$ &  6353 & 32 &2 &1 \\
$-32.4$ &  6784 & 29 &4 &4 & & $+  3.0$ &  6491 & 69 &2 &2 \\
$-31.1$ &  6763 & 35 &4 &4 & & $+  3.6$ &  6367 & 30 &2 &1 \\
$-29.8$ &  6796 & 47 &4 &4 & & $+  3.6$ &  6481 & 34 &2 &2 \\
$-28.5$ &  6784 &  0 &4 &4 & & $+  3.6$ &  6606 & 52 &2 &3 \\
$-27.1$ &  6783 &  3 &4 &4 & & $+  4.3$ &  6365 & 34 &2 &1 \\
$-25.8$ &  6784 & 12 &4 &4 & & $+  4.3$ &  6464 & 27 &2 &2 \\
$-24.5$ &  6729 &  0 &4 &4 & & $+  4.3$ &  6607 & 30 &2 &3 \\
$-22.8$ &  6721 &  0 &6 &4 & & $+  5.0$ &  6369 & 47 &2 &1 \\
$-20.9$ &  6799 & 10 &6 &4 & & $+  5.0$ &  6470 & 23 &2 &2 \\
$-18.9$ &  6736 &  0 &6 &4 & & $+  5.0$ &  6604 & 18 &2 &3 \\
$-17.2$ &  6790 & 99 &4 &4 & & $+  6.0$ &  6377 & 16 &4 &1 \\
$-16.1$ &  6749 & 22 &3 &4 & & $+  6.0$ &  6448 & 97 &4 &2 \\
$-15.1$ &  6755 & 35 &3 &4 & & $+  6.0$ &  6589 &  0 &4 &3 \\
$-14.1$ &  6738 &  6 &3 &4 & & $+  7.1$ &  6337 &  0 &3 &1 \\
$-13.1$ &  6735 &  9 &3 &2 & & $+  7.1$ &  6391 & 26 &3 &2 \\
$-13.1$ &  6789 & 20 &3 &1 & & $+  7.1$ &  6600 &  0 &3 &3 \\
$-12.1$ &  6766 & 50 &3 &4 & & $+  8.3$ &  6345 &  0 &2 &1 \\
$-11.3$ &  6799 & 21 &2 &4 & & $+  8.3$ &  6411 & 13 &2 &2 \\
$-10.6$ &  6802 & 32 &2 &4 & & $+  8.3$ &  6572 & 12 &2 &3 \\
$ -9.9$ &  6795 & 19 &2 &4 & & $+  9.1$ &  6348 & 18 &3 &1 \\
$ -9.3$ &  6786 & 28 &2 &4 & & $+  9.1$ &  6412 & 44 &3 &2 \\
$ -8.6$ &  6783 & 40 &2 &4 & & $+  9.1$ &  6569 & 14 &3 &3 \\
$ -7.9$ &  6757 & 39 &2 &2 & & $+ 10.1$ &  6351 & 17 &3 &1 \\
$ -7.9$ &  6827 &  0 &2 &1 & & $+ 10.1$ &  6381 & 26 &3 &2 \\
$ -7.3$ &  6747 & 73 &2 &2 & & $+ 10.1$ &  6565 & 16 &3 &3 \\
$ -7.3$ &  6814 &  0 &2 &1 & & $+ 11.1$ &  6362 & 10 &3 &1 \\
$ -6.6$ &  6741 & 72 &2 &2 & & $+ 11.1$ &  6371 & 39 &3 &2 \\
$ -6.6$ &  6818 &  0 &2 &1 & & $+ 11.1$ &  6549 & 23 &3 &3 \\
$ -6.0$ &  6740 & 55 &2 &2 & & $+ 12.4$ &  6370 &  3 &3 &1 \\
$ -5.3$ &  6727 & 69 &2 &2 & & $+ 12.4$ &  6378 & 37 &3 &2 \\
$ -5.3$ &  6810 &  0 &2 &1 & & $+ 12.4$ &  6552 & 14 &3 &3 \\
$ -4.6$ &  6731 & 57 &2 &2 & & $+ 13.4$ &  6369 & 27 &3 &4 \\
$ -4.6$ &  6815 & 15 &2 &1 & & $+ 14.4$ &  6376 & 20 &3 &4 \\
$ -4.0$ &  6711 & 41 &2 &2 & & $+ 15.4$ &  6350 & 23 &3 &4 \\
$ -4.0$ &  6809 &  6 &2 &1 & & $+ 16.4$ &  6345 & 18 &3 &4 \\
$ -3.3$ &  6685 & 41 &2 &2 & & $+ 17.4$ &  6338 & 28 &3 &4 \\
$ -3.3$ &  6810 &  0 &2 &1 & & $+ 18.4$ &  6337 & 24 &3 &4 \\
$ -2.7$ &  6662 & 46 &2 &2 & & $+ 19.4$ &  6324 & 22 &3 &4 \\
$ -2.0$ &  6641 & 47 &2 &2 & & $+ 20.4$ &  6315 & 18 &3 &4 \\
$ -1.3$ &  6613 & 84 &2 &2 & & $+ 21.4$ &  6323 & 22 &3 &4 \\
$ -0.7$ &  6505 &144 &2 &1 & & $+ 22.3$ &  6306 & 16 &3 &4 \\
$ -0.7$ &  6618 & 64 &2 &2 & & $+ 23.3$ &  6303 & 24 &3 &4 \\
$  0.0$ &  6422 & 73 &2 &1 & & $+ 24.3$ &  6310 & 26 &3 &4 \\
$  0.0$ &  6588 & 68 &2 &2 & & $+ 25.3$ &  6311 &  0 &3 &4 \\
$ +0.7$ &  6393 & 80 &2 &1 & & $+ 26.8$ &  6344 & 38 &6 &4 \\
$ +0.7$ &  6560 & 78 &2 &2 & & $+ 30.8$ &  6353 &  0 &6 &4 \\
 & & & & & & & & & & \\
\hline
\end{tabular}
\label{tabNII}
\end{scriptsize}
\begin{list}{}{}
\item[$^{\rm a}$] as in Table~1.
\end{list}
\end{flushleft}
\end{table}

\end{document}